\documentclass[12pt]{article}
\usepackage{pic04}
\usepackage{hyperref}
\usepackage{url}
\usepackage{graphicx}

\begin{document}

\title{\bf THE MiniBooNE EXPERIMENT}
\author{
Alexis Aguilar-Arevalo
 for the MiniBooNE collaboration \\
{\em Department of Physics, Columbia University,} \\
{\em 512 West 120th St., New York, NY, 10025, USA} }
\maketitle

\baselineskip=14.5pt
\begin{abstract}
MiniBooNE is an experiment designed to provide a 
definitive test for the 
$\overline{\nu}_\mu\rightarrow\overline{\nu}_e$ 
oscillations signal seen by LSND. Here, a brief 
summary of the MiniBooNE goals and strategies is 
presented, as well as some highlights of its 
current status.

\end{abstract}

\baselineskip=17pt

\section{Introduction}
The LSND  oscillations signal \cite{lsnd} (event excess of 
3.8 $\sigma$ significance) requires a %$\Delta m^2$
squared neutrino mass 
difference 
of the order of 0.3 - 3 eV$^2$. This is inconsistent 
with the three neutrino picture of the Standard Model (SM) 
when the oscillation evidence from solar and atmospheric 
neutrinos \cite{solatm} is taken into account. 
MiniBooNE's primary goal is to confirm or refute the 
oscillations interpretation of the LSND $\overline{\nu}_e$ excess.

\section{MiniBooNE neutrinos and detector overview}
The neutrino beam 
\footnote{For more details see Ref.\cite{runplan} and references 
therein}
is produced from the decay of 
charged mesons that are created when protons from the 
Fermilab
Booster (8.0 GeV kinetic energy) impact a 
%one interaction length
1.7 interaction length
Be target. The target is placed inside a focusing horn whose 
polarity can be set to run in either $\nu$ or $\overline\nu$ mode. 
Currently the experiment runs in $\nu$ mode, and the horn increases
the neutrino flux at the detector by a factor of $\sim$ 5.
The detector is a 12 m diameter sphere filled
with mineral oil (hydrocarbon chain). Its interior is 
covered 
\footnote{10\% photocathode coverage.}
by 1280 PMT's that gather the light produced by 
processes inside the tank. It is surrounded by an 
optically isolated veto region designed 
to tag cosmic rays with 240 PMT's . The transparent medium 
of the oil makes it primarily a Cerenkov detector, although 
an important component of scintillation light is present in 
any occuring process. The charge and time response of the 
tank PMT's is monitored continuously with a laser system. 
The energy scale of electron-type events is calibrated 
using the spectrum of Michel electrons. 
We use cosmic ray muons to determine the energy scale for 
muon-type events. The positions and angles of the incoming
cosmic rays are measured with a tracking hodoscope situated
above the detector and the stopping point is determined with
a system of 7 scintillating cubes located throughout the detector.

\section{Analysis status}
\begin{figure}[htbp]
  \centerline{\hbox{ \hspace{0.2cm}
    \includegraphics[width=5.5cm]{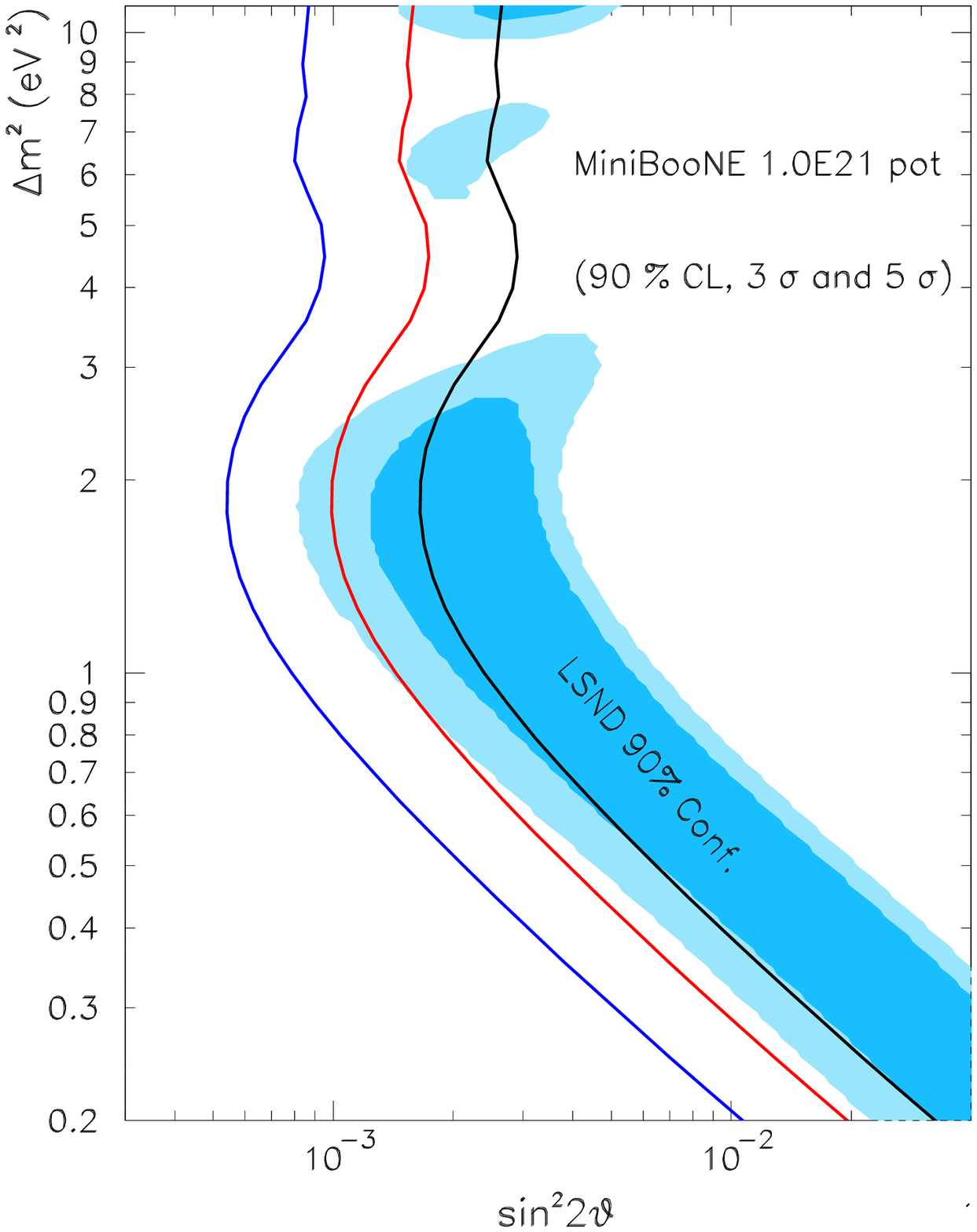}
    \hspace{1.3cm}
    \includegraphics[width=5.5cm]{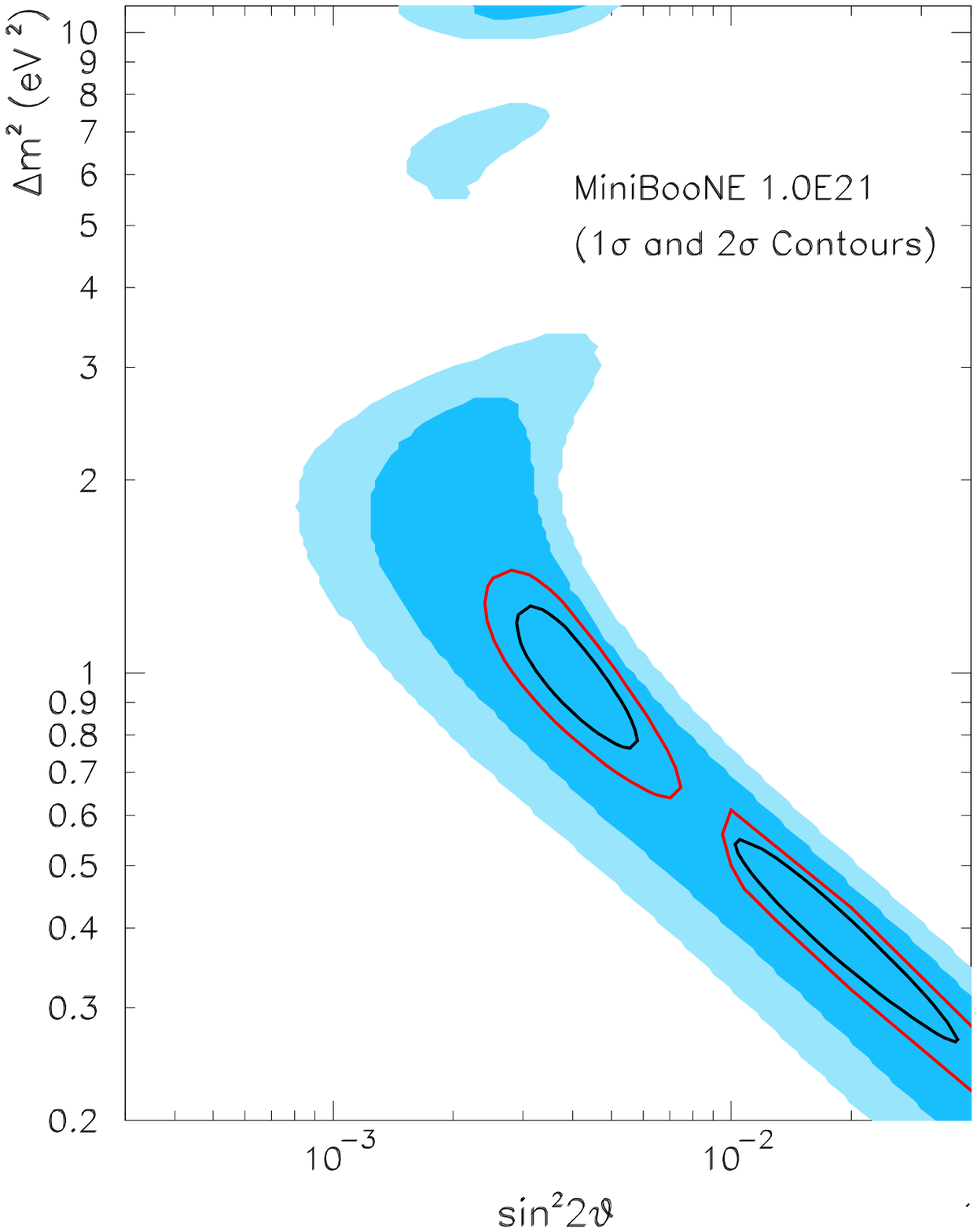}
    }
  }
 \caption{\it {\small
      Left: MiniBooNE $\nu_\mu\rightarrow\nu_e$ oscillation 
sensitiviy at 90\% C.L., 3$\sigma$, and 5$\sigma$ compared
to the LSND allowed region. Right: MiniBooNE 
$\nu_\mu\rightarrow\nu_e$ parameter measurement capability at
1$\sigma$ and 2$\sigma$ for high and low $\Delta m^2$.}
    \label{sensplot} }
\end{figure}

MiniBooNE is presently studying \footnote{For a detailed analysis 
update see Ref.\cite{neutrino04}. Plots shown in the poster 
are included therein.}  three processes that will be used to 
tune the detector Monte Carlo and the reconstruction algorithms: 
$\nu_\mu$ charged current quasi-elastic (CCQE) events which will 
be used to determine the incoming neutrino flux; neutral 
current (NC) $\pi^0$ production which will account for the 
largest $\nu_\mu$ misidentification background to the oscillation 
signal; and NC $\nu_\mu$ elastic scattering which can be  used 
to study the optical properties of the oil.  The 
$\nu_\mu\rightarrow\nu_e$ oscillation analysis is a blind 
analysis and, as such, we have not yet looked for $\nu_e$ 
events. However, recent studies \cite{runplan} show that 
with 1$\times 10^{21}$ protons on target (P.O.T.) approximately 
$300$ signal events are expected for LSND-like oscillations, 
as well as about $434$ misidentified $\nu_\mu$ interactions 
(dominated by $\sim 294$ NC $\pi^0$ production), and about 
$346$ intrinsic $\nu_e$ background events. Figure 
\ref{sensplot} shows the updated sensitivity and measurement 
capability for this number of protons.

\section{Conclusion}
The MiniBooNE experiment can definitively confirm or refute
the LSND oscillation signal with 1$\times 10^{21}$ P.O.T.
Currently, 30\% of this amount has been collected. 
The collaboration is presently working on the analysis of 
$\nu_\mu$ CCQE, NC $\pi^0$'s and NC elastic scattering.
The $\nu_\mu\rightarrow\nu_e$ analysis is 
expected to produce first results in 2005.

\section{Acknowledgements}
This work is supported by the National Science Foundation and the 
Department of Energy. The presenter of this poster was supported 
by NSF grant PHY-98-13383.

\end{document}